\documentclass[review]{elsarticle}

\usepackage{lineno,hyperref}
\modulolinenumbers[5]

\journal{Food Webs}

\biboptions{authoryear}

\usepackage{amsmath} 

\begin{document}

\begin{frontmatter}

\title{Robustness of soil ecosystems under different regimes of management}

\author[mymainaddress,mysecondaryaddress]{Letizia Stella Di Mauro\corref{mycorrespondingauthor}}
\cortext[mycorrespondingauthor]{Corresponding Author}
\ead{letizia.dimauro@ct.infn.it}

\author[mymainaddress2]{Christian Mulder}

\author[mymainaddress2]{Erminia Conti}

\author[mymainaddress,mysecondaryaddress]{Alessandro Pluchino}

\address[mymainaddress]{Department of Physics and Astronomy "Ettore Majorana", University of Catania, Via S. Sofia 64, 95123 Catania, Italy}
\address[mysecondaryaddress]{INFN Unit of Catania, Via S. Sofia 64, 95123 Catania, Italy}
\address[mymainaddress2]{Department of Biological, Geological and Environmmental Sciences, University of Catania, Via Androne 81, 95124 Catania, Italy}

\begin{abstract}
In this study three soil ecosystems, that differ in the type of management, have been compared in the attempt to understand if and how anthropogenic action affects them. The structure of the corresponding food webs was analyzed and their robustness was calculated through the use of a dynamic model.
With regard to the topology, it has been found that the structure of all three networks is small world. Furthermore, all three networks have a disassortative nature as expected for foodwebs. The values of the clustering coefficient, of the connectance and of the complexity, together with the calculation of the robustness suggest that the ecosystem related to a fallowed pastures with low pressure management is more robust than two ecosystems related to organic farms subject to middle intensity management.
If the evidence suggested by this study was confirmed by further studies, the robustness shown by the networks could be useful for evaluating, from an ecological point of view, the sustainability of the agricultural practices to which the ecosystem is subject.
\end{abstract}


\end{frontmatter}
\thispagestyle{plain}


\section{Introduction}

One of the most controversial debates in the context of food networks is the relationship between stability and complexity of an ecological system.\\ The predominant idea that the stability of ecological communities would increase with their complexity (\cite{Elton, MacArthur}), was questioned by \cite{May} who, using methods related to dynamic models,  came to the conclusion that the stability of an ecosystem decreases with increasing number of species and interactions.
 May's criterion suggests that a community remains stable if a decrease in connectance $C$ is accompanied by an increase in diversity $S$, so that $SC$ remains a constant quantity (\cite{Pascual, Mulder-et-al-2006}). Many studies, among which \cite{Cohen-Newman} and \cite{Cohen-et-al}, at first seemed to corroborate May's hypothesis and found approximately constant values of linkage density $SC\simeq 2$. Contrarywise new research, based on more detailed empirical data, found a higher degree of interaction between species, $SC\simeq 10$ (\cite{Martinez}), and a positive relationship between $C$ and $S$ contrary to what predicted by the May criterion (\cite{Sugihara}).\\
As concerns the topology, a study carried out by \cite{Montoya-Sole} suggests that food webs show a small-world and scale-free structure, showing high values of the clustering coefficient and a power-law degree distribution. Instead other studies have highlighted a deviation from the small-world and scale-free topology founding that most food webs display low clustering coefficients, similar to random expectations, and less skewed exponential and uniform distributions especially, in the case of food webs with high connectance values (\cite{Dunne-et-al-2002a, Camacho}).
On the contrary, as regards the degree correlation, the disassortative nature of trophic networks seems to have been ascertained: nodes with many links are mostly connected with nodes with a low number of links (\cite{Newman-MEJ-2002, Newman-MEJ-2003, Stouffer}).\\
Research on trophic networks concerns topics of great ecological interest, like their robustness in response to external perturbations that depends on its structure. \cite{Sole-Montoya} found that food webs are more vulnerable to targeted attacks to hubs than to random attacks, characteristic generally found in scale-free networks (\cite{Strogatz, Barabasi}). Other studies found that even without highly skewed degree distributions, food webs are much more robust to random loss of species than to loss of highly connected species, suggesting that any substantial skewness in degree distribution will tend to alter the response of a network to different kinds of node loss (\cite{Dunne-et-al-2002b, Dunne-et-al-2004}). Moreover \cite{Dunne-et-al-2002b} also found that the robustness of the food web increases with increasing connectance and this result applies both for targeted removals of  hyper-connected species and for random removals of nodes.\\
The aim of the study described in this paper is to analyze the topology and the robustness of three trophic networks corresponding to soil ecosystems under different regimes of management, highlighting similarities and differences. Through the use of a model whose dynamics is dictated by an extension of the Lotka-Volterra equations (\cite{Conti}), we study what may be the consequences of artificial perturbations induced in the system. Unlike previous studies, that are based on a purely topological network-structure analysis, this study derives the robustness of the food webs through a dynamical analysis. This represents an upgrade considering that the structure of a given network has a strong impact on the outcomes of dynamics, as highlighted in \cite{Pimm, McCann, Hastings, Jordan}. In this regard, \cite{Dunne-2006} stated that ``\textit{the dynamics of species in complex ecosystems are more tightly connected than conventionally thought, which has profound implications for the impact and spread of perturbations}".\\
Several studies have investigated the different structural characteristics shown by food webs representative of different types of habitats and environments (\cite{Briand-1983, Briand-Cohen, Chase, Cohen-1994, Link, Dunne-et-al-2004}). 
Similarly, the present study is useful in highlighting structural differences between food networks representative of agricultural fields subject to different types of management, differences that may affect the dynamics and robustness of these networks.
Land-management practices and environmental changes affect belowground communities, influencing the overall stability and productivity of the food webs  (\cite{Wall, Clay, Powell}).
In this respect this study aims also to suggest a possible link between agricultural practices and the robustness of ecosystems subject to these types of management. Understand if and how anthropogenic action affects soil ecosystems can be helpful in establishing the eco-sustainability of certain agricultural methods, promoting the ecological complexity and robustness of soil biodiversity.

\section{Material and methods}
\subsection{Data sampling and construction of the food webs}
\begin{figure}[tbp]
\centering
\includegraphics[width=0.9\columnwidth]{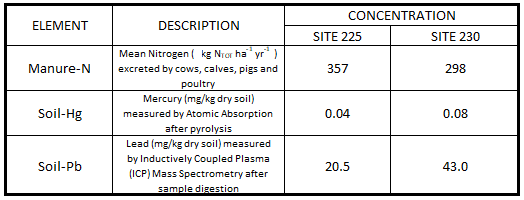}
\caption{Concentration values of nitrogen from manure, mercury and lead in the soil of sites 225 and 230. }
\label{concentration}
\end{figure}

\begin{figure}[tbp]
\centering
\includegraphics[width=0.8\columnwidth]{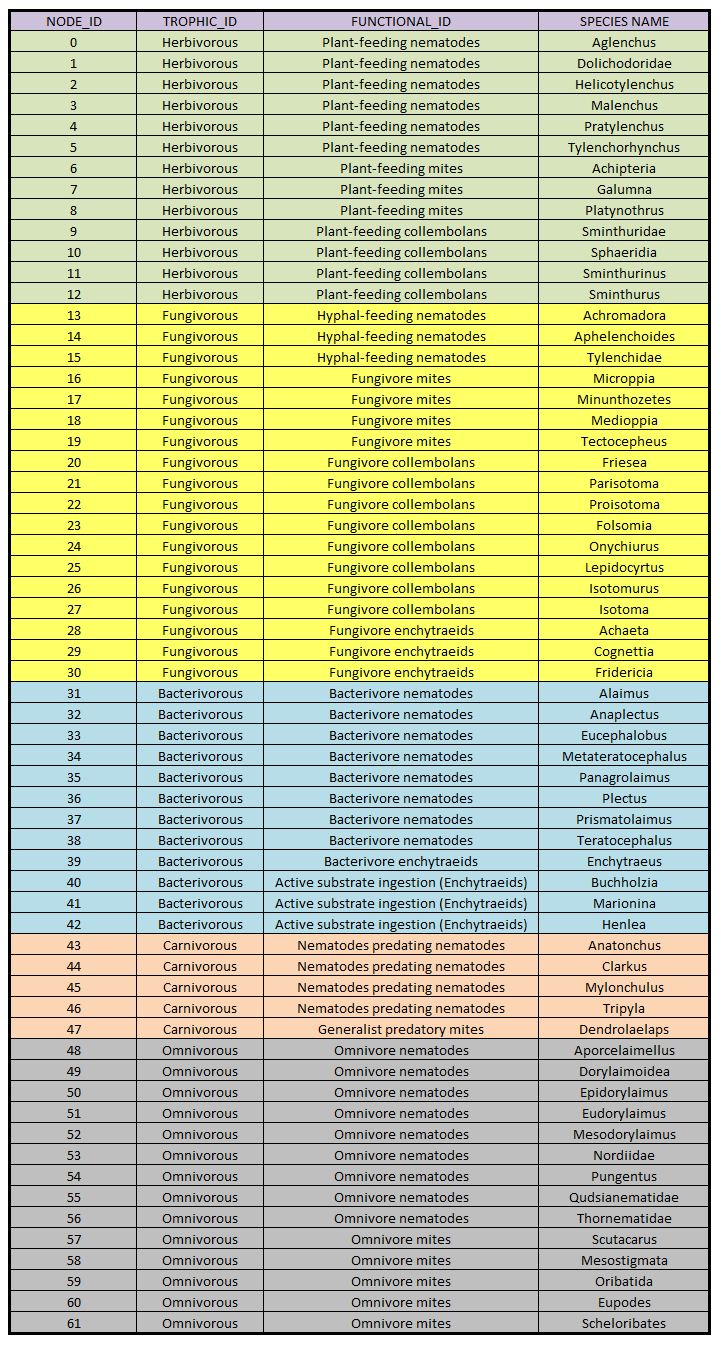}
\caption{Trophic ID, functional ID and name of the taxa (mostly families or genera, here after called “species”) corresponding to the nodes of the network shown in Figure~\ref{Networks} (a) (site 247).}
\label{tabella_247}
\end{figure}

\begin{figure}[tbp]
\centering
\includegraphics[width=0.8\columnwidth]{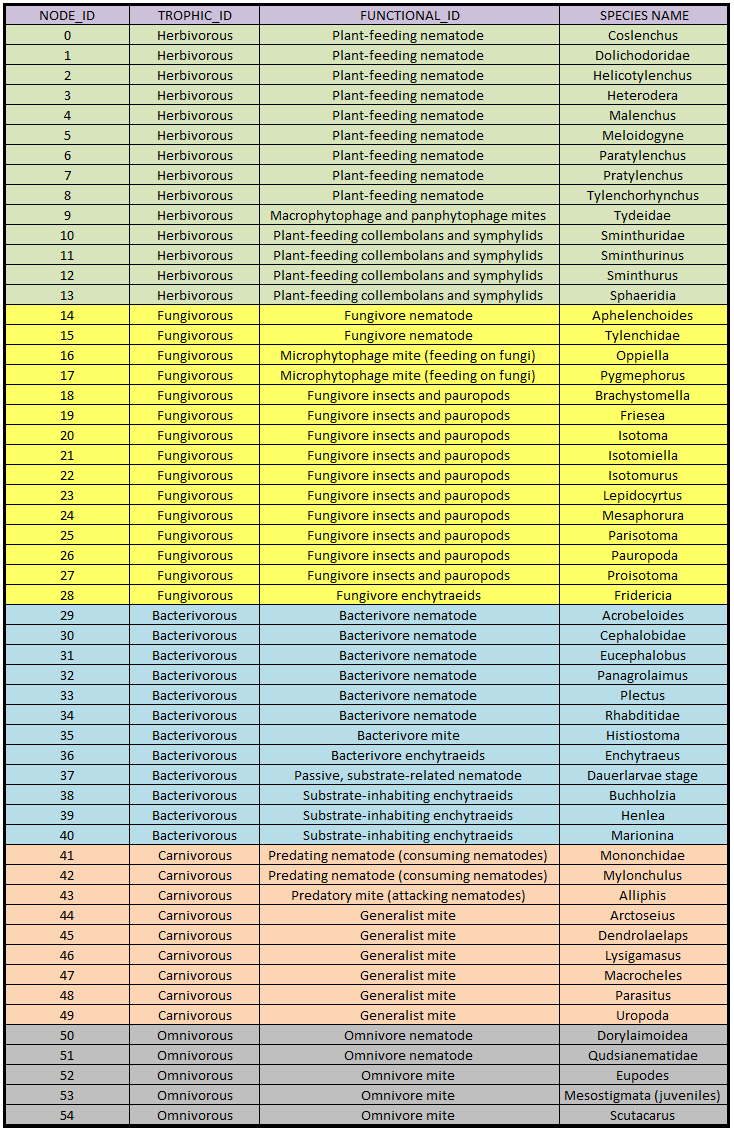}
\caption{Trophic ID, functional ID and name of the taxa (mostly families or genera, here after called “species”) corresponding to the nodes of the network shown in Figure~\ref{Networks} (b) (site 225).}
\label{tabella_225}
\end{figure}

\begin{figure}[tbp]
\centering
\includegraphics[width=0.8\columnwidth]{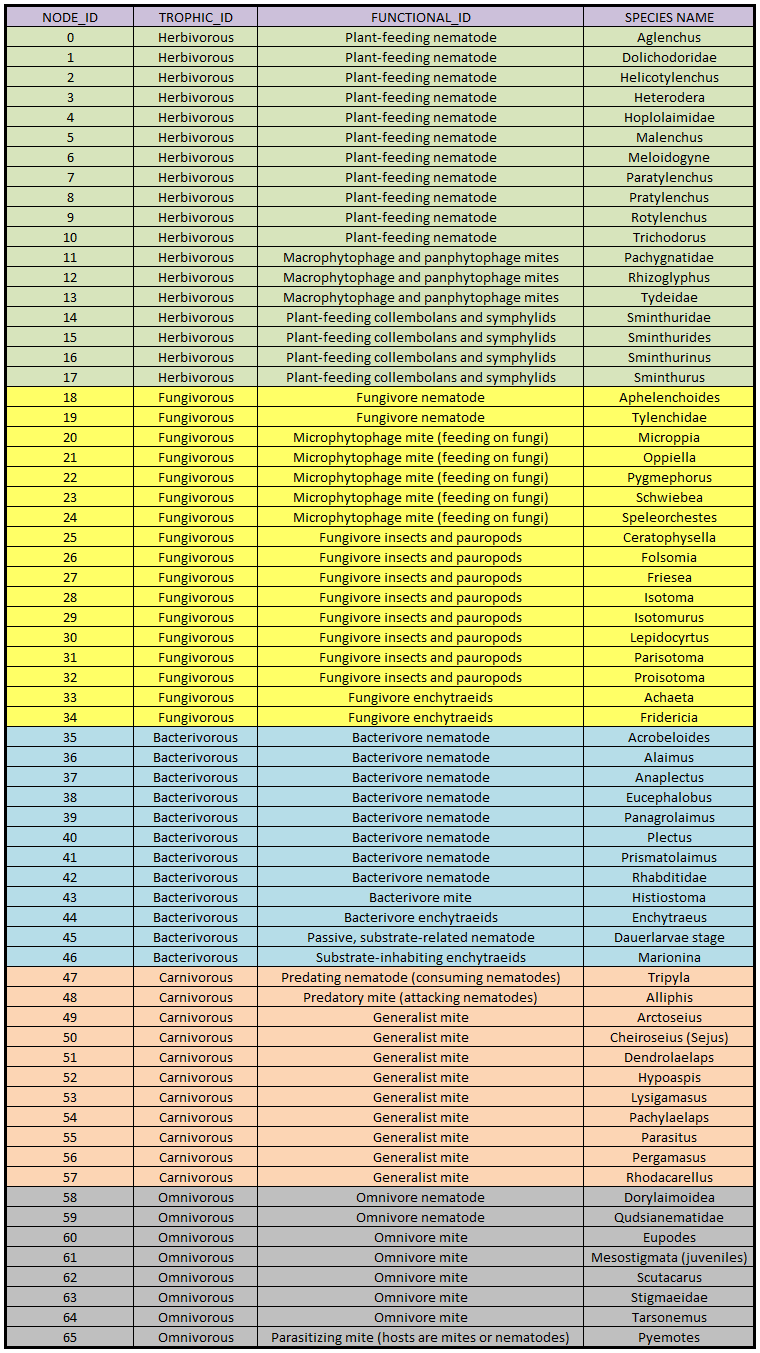}
\caption{Trophic ID, functional ID and name of the taxa (mostly families or genera, here after called “species”) corresponding to the nodes of the network shown in Figure~\ref{Networks} (c) (site 230).}
\label{tabella_230}
\end{figure}

\begin{figure}[tbp]
\centering
\includegraphics[width=1\columnwidth]{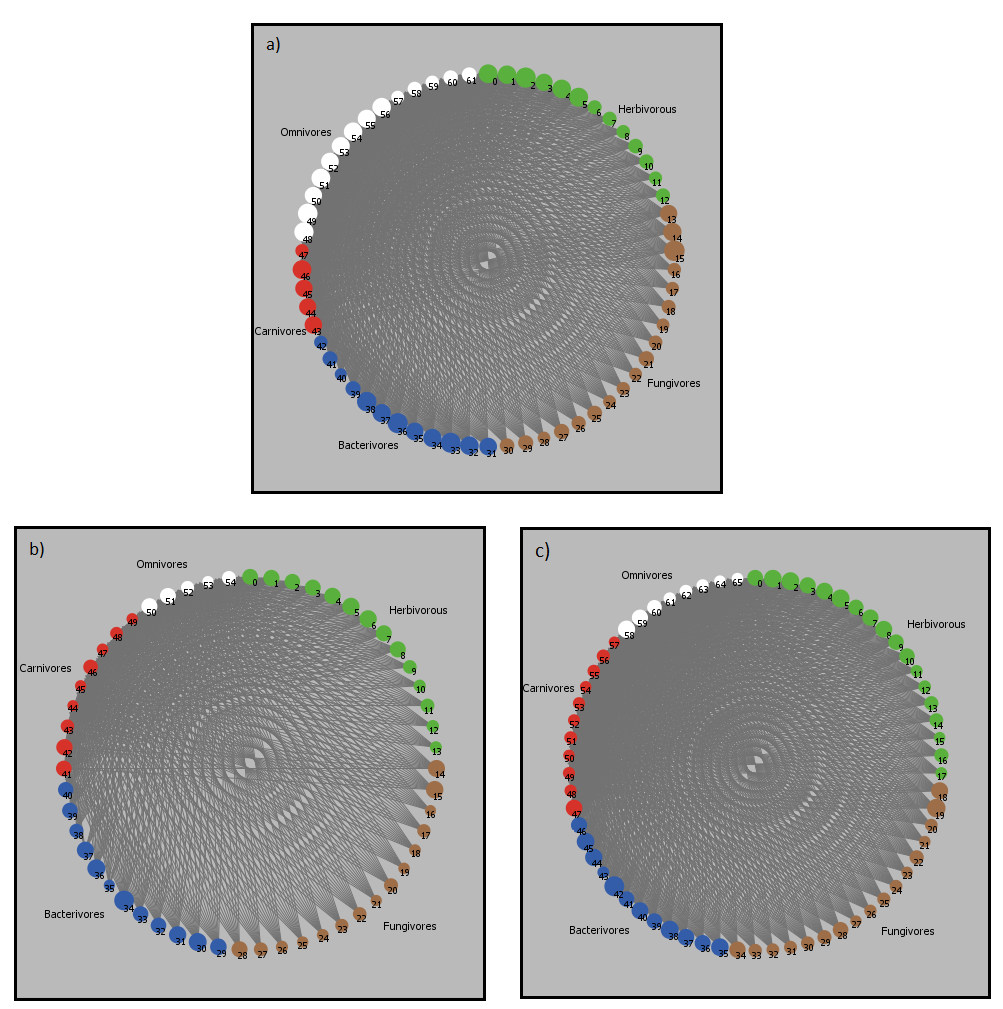}
\caption{A sketch of the food web networks for site 247 (a), site 225 (b) and site 230 (c). Nodes/species are organized in groups of different colors (see Figures ~\ref{tabella_247} ~\ref{tabella_225} ~\ref{tabella_230}) and placed in a circular layout. Directed links, between the organisms belonging to trophic level 1 (primary consumers, being they herbivorous, fungivorous or bacterivorous invertebrates) and the organisms belonging to trophic level 2 (secondary consumers, either carnivorous or omnivorous invertebrates), represent the prey/predator connections, going from the prey node to the predator node.}
\label{Networks}
\end{figure}

The data used in this study derived from sampling and monitoring activities performed in the framework of the Dutch Soil Quality Network. Thanks to this survey, we have available data on the taxonomy, abundance, body size, and general feeding habits of soil invertebrates at 135 sites in the Netherlands.\\
This study focused only on three sampling sites and can be considered as a preliminary study for further analysis to be carried out on the remaining available sites.
The first ecosystem considered (site 247) is a field with low pressure management, in particular, a fallowed pasture, described in \cite{Conti}. According to the Eltonian rule (\cite{Elton-1927}),  we can consider this ecosystem as a reference network subject to low anthropic disturbance (\cite{Mulder-Elser-2009, Conti}).\\
The other two ecosystems (sites 225 and 230) are organic farms certified by the Agricultural Economics Research Institute of the Netherlands (LEI). These meet all the legal requirements for this type of agriculture (using compost/farmyard manure and no biocides, averaging 1.7 livestock units per hectare) and are periodically monitored (\cite{Cohen-Mulder-2014}).\\
Furthermore, the two organic farms differ from each other in the concentration of some elements present in their soil.
The first of these two ecosystems (site 225) is a field with high concentrations of nitrogen from manure and this suggests that it is an area subject to grazing (organic farming also includes the possibility of pasture).
The second ecosystem (site 230) was instead selected for its high concentrations of heavy metals such as lead and mercury which might suggest the presence of a nearby factory functioning at present or in the past. The concentration values of these significant elements in the two sites are shown in Figure~\ref{concentration}.\\
Regarding soil sampling, three replicate samples of about 5 $m^2$ from the upper 10 cm of soil for the fauna were taken. Bulk samples of 50 soil cores (diameter 2.3 cm) were used to extract the microfauna and two soil cores (diameter 5.8 cm) were used to extract the mesofauna. All sampling protocols were extensively described in \cite{Mulder-et-al-2011}.\\
In Figures ~\ref{tabella_247} ~\ref{tabella_225} ~\ref{tabella_230}, the list of the species found in the three sampling sites is shown. For these species we know the abundance $X_i$, the body mass $M_i$, the biomass $B_i$ (given by $B_i=X_i M_i$) and the value of the growth rate $r_i$ in condition of mutual interaction, with $i= 0,1, ... ,n$ ($n$ is the number of species detected in the ecosystem). Within each guild, we derived the $r_i$ values for the functional groups shown in Figures ~\ref{tabella_247} ~\ref{tabella_225} ~\ref{tabella_230} from \cite{Moore-et-al-1993,  De-Ruiter-et-al-1995}. All identified soil invertebrates fell into five guilds and the independent trophic links among guilds (from any resource to its consumer) were inferred from \cite{Mulder-Elser-2009}. Using the data obtained from the sampling, we have built three direct and unweighted food webs corresponding to the three ecosystems of soil organisms. In Figure ~\ref{Networks} the representative networks of the three ecosystems are showed, with the numbered nodes/species placed in a circular layout, where each group of species is distinguished by a different color (as also explained in detail in Figures ~\ref{tabella_247} ~\ref{tabella_225} ~\ref{tabella_230}). The size of each node is proportional to the base-10 logarithm of the abundance of the corresponding species  ($Log X$).\\

\subsection{Simulation of dynamics within the ecosystems}
The dynamics within the ecosystem was simulated using the model and applying the same methodology described in \cite{Conti}. That specific model has recently been implemented only for network 247, simulating some scenarios that analyze the dynamics of the system after the removal of some groups of species. Simulations have been carried out within NetLogo, a multi-agent fully programmable modeling environment particularly suitable for the simulation of complex systems (\cite{Wilensky}). In the model we combine the Lotka-Volterra model (\cite{Lotka, Volterra-1926, Volterra-1939}) with the logistic equation (\cite{Verhulst}) in order to take into account intra-specific and inter-specific competition.
The dynamical equations describing the variation in the abundance of the species, are given by 
\begin{linenomath}
\begin{equation}
\label{eq:dinamics}
\frac{dX_i}{dt}=\left[r_{i0}+ \alpha \sum_j{A_{ij} X_j}\right]X_i\left(1-\frac{X_i}{K_i}\right) \; (i = 1, 2, ..., n)
\end{equation}
\end{linenomath}
where $r_{i0}$ is the growth rate of the i-th species in absence of interaction and $K_i$ its carrying capacity.
Coefficients $A_{ij}$ weigh the food interaction among pairs of species and depend on the biomass of the prey species. In particular, $A_{ij}>0$ when species $i$ preys species $j$ and $A_{ij}<0$ when species $j$ preys species $i$.
The parameter $\alpha$ is a coupling coefficient that can be considered as a measure of the interaction strength of a given species within the rest of the food web.
For details on how the values of $A_{ij}$, $r_{i0}$, $K_i$ and $\alpha$ were obtained, refer to the methodology described in
\cite{Conti}. 
Here we just point out that the formula has the same form of a logistic equation in which instead of the growth rate without interaction $r_{i0}$, the term inside the square brackets in equation \ref{eq:dinamics} is used. This term is the growth rate $r_i$ in case of interaction between species and considers, in addition to the intra-specific interaction an inter-specific interaction, i.e. the effect due to predation:
\begin{linenomath}
\begin{equation}
\label{eq:growth_rate}
r_i=r_{i0}+ \alpha \sum_j{ A_{ij} X_j}
\end{equation}
\end{linenomath}
All the simulations were done by choosing the initial abundance of the species in the interval:
\begin{linenomath}
\begin{equation}
\label{eq:initial-conditions}
X_i(0) \in \left[K_i- \frac{K_i}{2};K_i\right]
\end{equation}
\end{linenomath}
so that they cannot exceed their carriyng capacity.
For each considered food web, starting from the initial conditions \ref{eq:initial-conditions}, at each time step the populations $X_i(t)$ of all species are updated by numerically integrating equation \ref{eq:dinamics} and all the species reach the steady state established by their carrying capacity. Any removal of one or more species will induce a disturbance within the system which could lead to a variation in the abundance or, possibly, to the extinction of some species.\\
In order to quantify structural changes and to compare one single simulation to others, we use the \textit{Alteration Index (AI)}  introduced in \cite{Conti} and defined as:
\begin{linenomath}
\begin{equation}
\label{eq:ai}
AI = \sum_k \frac{|X_{sk}-X_{fk}|}{X_{sk}}=\sum_k \frac{|\Delta X_k|}{X_{sk}}
\end{equation}
\end{linenomath}
where $X_{sk}$ and $X_{fk}$ are, respectively, the abundance of species k-th calculated after 100 time steps, i.e. in the steady state, and the abundance of the same species calculated at the end of the simulation. In other words, $AI$ considers the sum of the absolute variations in abundance that the species undergo due to the forced removal of some other species, normalized with respect to their abundance in the steady state. It is therefore a measure of the alteration of the ecosystem due to the introduced perturbation.

\section{Theory and calculation}

\subsection{Biodiversity, Connectance and Complexity}
In the case of trophic networks, the number of nodes and therefore of species, defines the biodiversity $S$ of the ecosystem.
By calling $L$ the number of trophic links between the species, each graph is distinguished by the value of the connectance $C$, which gives the ratio between the number of connections actually present on the possible ones. Therefore we have $C =L/S^2$ in the event that loops (i.e. connections of a node with itself) are taken into consideration, that, in the case of food webs, is equivalent to consider the phenomenon of cannibalism. Otherwise, if we do not consider the presence of loops, we have  $C = L/S (S - 1 )$. Thereby connectance gives a measure of the probability that two species interact with each other within a graph.
The complexity $c$ of a network, and in particular in this case of an ecosystem, is closely connected to the concept of connectance. Complexity is in fact defined as the product of the connectance $C$ for the biodiversity $S$ of the ecosystem and corresponds to the linkage density $L/S$ of the web: $c=CS=L/S$.

\subsection{Network topology}
The topology of a network is strongly determined by two characteristic quantities: the average path length $\langle d \rangle$ and the average clustering coefficient  $\langle C \rangle$.
The average path length $\langle d \rangle$ is given by the average of the distances $d_{ij}$ between all pairs of nodes in the network. For a direct network this quantity is given by
\begin{linenomath}
\begin{equation}
 \langle d \rangle =\frac{1}{S (S-1)} \sum_{\substack {i,j=1,S \\ i \neq j}} d_{ij} 
\end{equation}
\end{linenomath}
\\
The average clustering coefficient $\langle C \rangle$ gives an idea of how strong the aggregation between the nodes is. For a single node $i$ with degree $k_i$, the local clustering coefficient is given by
\begin{linenomath}
\begin{equation}
 C_i =\frac{2L_i}{k_i (k_i -1)}
\end{equation}
\end{linenomath}
where $L_i$ represents the number of links that connect the $k_i$ neighbor nodes of node $i$.
Basically $C_i$ gives the probability that two neighbors of a node are in turn connected and it is a quantity between 0 and 1.
The degree of clustering of the entire network is determined by the average clustering coefficient
\begin{linenomath}
\begin{equation}
 \langle C \rangle =\frac{1}{S} \sum_{i=1}^S C_i
\end{equation}
\end{linenomath}
In random graphs the average path length and the average clustering coefficient are both low. \cite{Watts-Strogatz} proposed the Small World graph that can be built from a regular graph by replacing some of its links with random links. This type of graph presents a low average path length and, as opposed to the random graph, a high average clustering coefficient, features that are simultaneously present in the structure of many real networks.
To say that a network is a small world one, it is necessary to ensure that it has a low value of the average path length and at the same time a high value of the clustering coefficient compared to that of a random network comparable to it, that is, with the same number of links and nodes.\\
In egalitarian small world networks all nodes have about the same number of links. However, many real networks do not follow a Poisson degree distribution and are aristocratic, i.e. consisting of a majority of nodes with few links and a minority of hyper-connected nodes, called hubs. Examples of real networks of this type are the World Wide Web (\cite{Albert-et-al}), the Internet (\cite{Faloutsos}), the network of scientific collaborations (\cite{Newman-2001}), the network of sexual contacts (\cite{Liljeros-et-al}), the protein networks (\cite{Jeong-et-al-2001}) or the metabolic networks (\cite{Jeong-et-al-2000}). Since they do not have a typical scale, they are called scale-free networks. Their degree distribution follows a power law:
\begin{linenomath}
\begin{equation}
p_k \sim k^{-\gamma}
\end{equation}
\end{linenomath}
The different nature of the systems which, if described in terms of complex networks, show the scale-free property make it an almost universal feature.\\
Many studies questioned the structure presented by trophic networks. In particular, \cite{Montoya-Sole} compared the properties of real food networks with those obtained for random networks with the same number of links and found that the average path length is very similar and very short, but the clustering coefficient is much greater for real food webs compared to random ones. This, as already mentioned, is a characteristic aspect of small-world behavior. In addition, they obtained a strongly non-Poissonian link distribution $P(k)$ which seems to follow a power law.
\cite{Camacho} contradicted these results by stating that the clustering coefficient of real food chains is lower than that observed in small-world networks and therefore more similar to that of a random network. Furthermore, according to them, link distribution does not appear to be scale-free. However, they found that, when link distribution is normalized for link density $L/S$, it shows a universal functional shape given by an exponential decay instead of by a power law one. Also the clustering coefficient and the average path length seem, according to \cite{Camacho}, to follow a universal functional form that scales with the density of the links.

\subsection{Network Robustness}
Some natural and social systems show a great capacity in maintaining basic functions despite the failure or lack of some of its components. The percolation theory studies the robustness of networks by assessing the impact of removing nodes or, alternatively, links. Robustness is inferred from the percentage of nodes that must be removed to completely break up the system.
In percolation theory, to get an idea of the degree of disintegration of the system, the measure of the largest component of the network (giant component) is considered. Alternatively, \cite{Dunne-et-al-2002b} define the robustness of the food web as the fraction of primary species removed which induces a total loss of at least 50 \% of the species (primary and secondary extinctions). 
In this study robustness calculation will be carried out according to this definition.
The removal or failure of one node is not independent of the others because the activity of each node depends on the activity of its neighboring nodes. Therefore, cascade failures could be observed in which the failure of a node induces the failure of the nodes connected to it, as in the domino effect in which a local variation propagates throughout the whole system. For this reason it is important to perform the calculations of the robustness of the network, not only from a structural point of view, but simulating the dynamics resulting from the removal of the nodes, as is done in the present study.\\
The topology of the graph, and in particular the presence of hubs in scale-free networks, strongly influences its resistance to external attacks, determining the robustness of the system. The greater robustness of scale-free networks compared to random ones, in response to the random removal of links, is due to the presence of hubs. Being random, removal will be much more likely to involve nodes that have a low degree because these are much more numerous than hubs. On the contrary, hubs will be removed with an extremely lower probability and this is what allows the network to remain intact. The question is different in the case of targeted attacks on the system rather than random removals. Assuming to know in detail the topology of the network, attacks aimed at removing nodes with a high degree can be perpetrated. The removal of even a small fraction of  hubs is sufficient to disrupt a scale-free network. The systems with a scale-free networks are therefore very tolerant of random errors or failures, but very vulnerable to targeted attacks.\\
\cite{Sole-Montoya} studied the response of some food chains by simulating the loss of nodes and looking at the level of secondary extinctions. They came to the conclusion that the removal of highly connected species causes a very high rate of secondary extinctions compared to a random removal of species. It would therefore seem that food networks are more vulnerable to targeted attacks to hubs than to random attacks, characteristic generally found in scale-free networks (\cite{Strogatz, Barabasi}). 
As mentioned previously, however, trophic networks do not seem to have a scale-free structure. According to \cite{Dunne-et-al-2002b, Dunne-et-al-2004}, the degree distribution still being fat tailed, even if not properly with a power law slope, alters the response to targeted and random removals so that the first modality is more effective than the second, similarly to what happens in  scale-free networks.\\
As specified by \cite{Dunne-2006}, in the specific case of food webs, the removal of the most interconnected species is not always the best strategy to carry out targeted attacks affecting the integrity of the ecosystem. In particular, \cite{Allesina-Bodini} have shown that the most important species for the integrity of the system are the dominant species, that is those that pass energy to other species along the food chain. It is precisely the removal of these species that causes a greater number of secondary extinctions. The dominant species, although probably having a high out-degree value, are not necessarily the ones most interconnected if ingoing links are also considered. For this reason, in this study, the criterion for the removal of species in targeted attacks is the elimination of those nodes with a high closeness centrality value. The closeness centrality of a node is defined as the inverse of the average of its distances to all other nodes. It measures how close a node is to the others and quantifies how rapidly an effect that generates from that species can spread in the food web (\cite{Rocchi-et-al}).
It should be noted that, in the food webs studied here, the species with the highest closeness centrality value are also the dominant ones, that is, the most abundant ones and those with a high out-degree value.

\subsection{Degree Correlation}
Degree correlation is a property of graphs that regard the tendency of nodes to connect with other nodes of similar or completely different degrees. Based on this characteristic, three types of networks are distinguished.
In neutral networks nodes link to each other randomly, so the number of links between the hubs coincides with what predected by chance. In assortative networks  nodes with comparable degree tend to link each other: small-degree nodes to small-degree nodes and hubs to hubs. Finally, in disassortative networks the hubs avoid each other, linking instead to small-degree nodes.
A simple way to quantify the degree correlation makes use of the degree correlation function (\cite{Barabasi}). For each node $i$ we can measure the average degree of its neighbors:
\begin{linenomath}
\begin{equation}
k_{nn}(k_i)= \frac {1}{k_i} \sum_{j=1}^SA_{ij}k_j
\label{eq:degree-correlation}
\end{equation}
\end{linenomath}
The degree correlation function calculates equation (\ref{eq:degree-correlation}) for all nodes with degree $k$:
\begin{linenomath}
\begin{equation}
k_{nn}(k)=  \sum_{k'} k'P(k'|k)
\label{eq:degree-correlation-2}
\end{equation}
\end{linenomath}
where $P(k'|k)$ is the conditional probability that by following a link of a degree-$k$ node, we reach a degree-$k'$ node (\cite{Barabasi}). Therefore, $k_{nn}(k)$ is the average degree of the neighbors of all degree-$k$ nodes.
If we approximate the degree correlation function with
\begin{linenomath}
\begin{equation}
k_{nn}(k)=  ak^{\mu}
\label{eq:degree-correlation-3}
\end{equation}
\end{linenomath}
the nature of the degree correlation is determined by the sign of the correlation exponent $\mu$: positive for assoratative neworks, negative for disassotative networks and almost zero for neutral networks.\\
There are many studies investigating the properties of networks that derive from their degree correlation, among which \cite{Murakami, D'Agostino, Tanizawa, Thedchanamoorthy}.
According to these studies, assortative networks have the capacity to be more robust against targeted attacks, while disassotative networks have greater efficiency in the transport of information. This would explain why communication-oriented networks, i.e. networks whose primary function is the exchange of information, have evolved a disassortative structure. In assortative networks, hub removal in targeted attack causes less damage because the hubs form a core group, hence many of them are redundant. Hub removal is more damaging in disassortative networks, as in these the hubs connect to many small-degree nodes, which fall off the network once a hub is deleted. Real world networks display assortative hubs in some instances, particularly when high robustness to targeted attacks is a necessity (\cite{Thedchanamoorthy}).
The disassortative nature of trophic networks (\cite{Newman-MEJ-2002, Newman-MEJ-2003, Stouffer}) could explain their weakness towards targeted attacks to the most interconnected nodes, more than their degree distribution.

\section{Results and Discussion}
\subsection{Topology of the three food webs}
\begin{figure}[tbp]
\centering
\includegraphics[width=0.8\columnwidth]{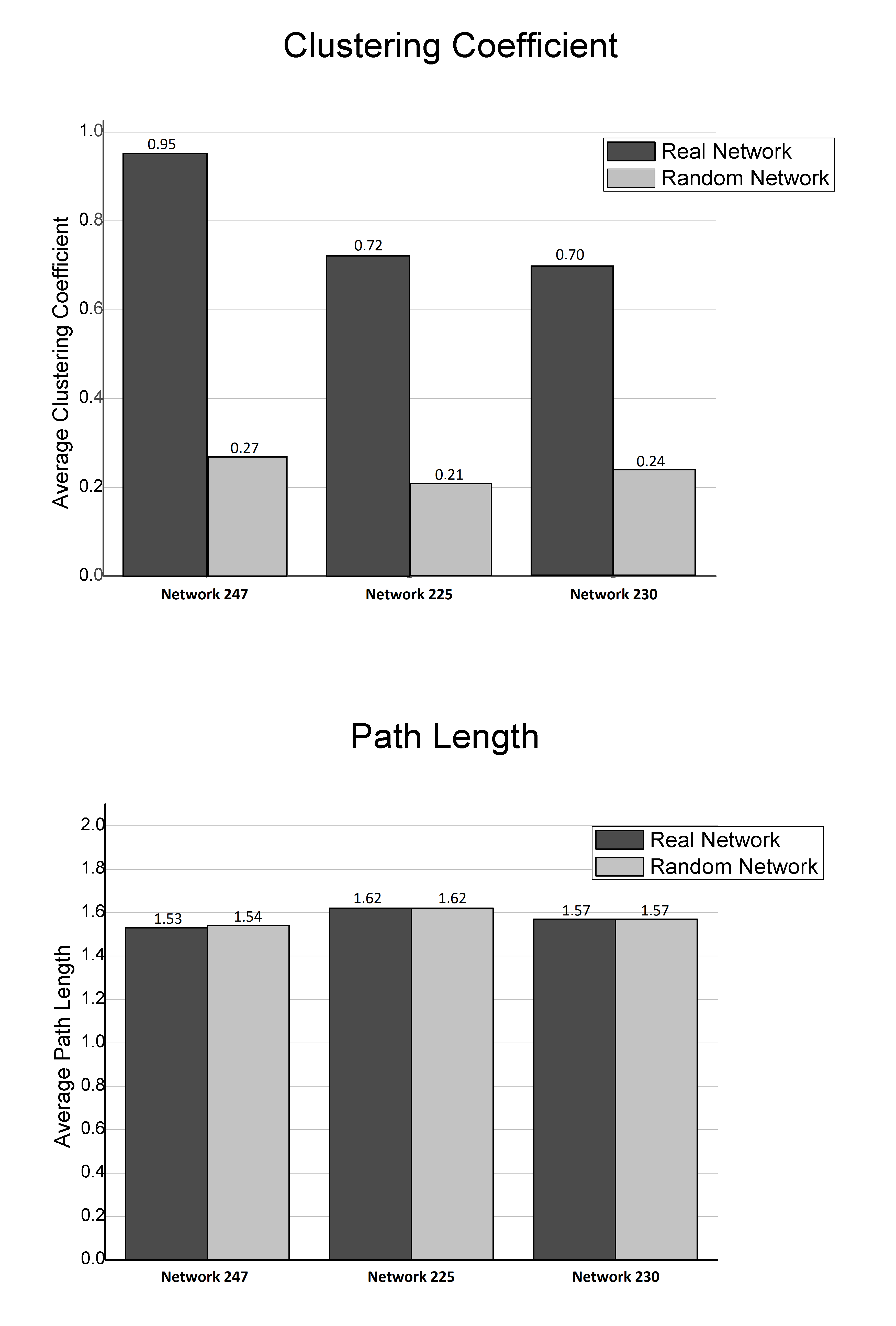}
\caption{Average Clustering Coefficient and Average Path Length. The dark grey columns represented the real networks and the light grey columns the average results for 10 comparable random networks.}
\label{small-world}
\end{figure}

To check if the three food networks considered in this study have a small world structure we calculate their average path length and their average clustering coefficient and we compare these results with the average values of the same quantities obtained from 10 random networks with the same number of nodes and the same average number of links. Since the considered direct networks are disconnected, for the calculation of the average path length it was preferred to take into account the corresponding indirect graph. Results are shown in Figure~\ref{small-world}.
The average clustering coefficient of real networks is much greater than that obtained for random networks. Instead the average path length assumes a value similar, if not equal, to that of the random counterpart. These results are in line with those obtained by  \cite{Montoya-Sole} for other types of food webs. The low values of  the average path length and the great difference between the average clustering coefficient of real and random networks can be considered a confirmation of the small world nature of the three networks analyzed. \\
By comparing the values of the average clustering coefficient of the three real networks, we observe that networks 225 and 230, both organic farms, have very similar and, evidently, lower clustering coefficients than network 247, an uncultivated land. A higher clustering coefficient could indicate a greater robustness of the system, since the presence of clusters within the network guarantees the presence of alternative routes in the event of disappearance of nodes.\\
For all three networks the value of the connectance, given by $C = L/S (S - 1 )$ (as the phenomenon of cannibalism was not taken into consideration), was calculated. The connectance values are 0.27, 0.21 and 0.24 respectively for networks 247, 225 and 230. According to \cite{Dunne-et-al-2002b}, low connectance values may reveal a lower robustness of the network. Thus, also in this respect, network 247 seems to be the most robust.\\

\begin{figure}[tbp]
\centering
\includegraphics[width=0.6\columnwidth]{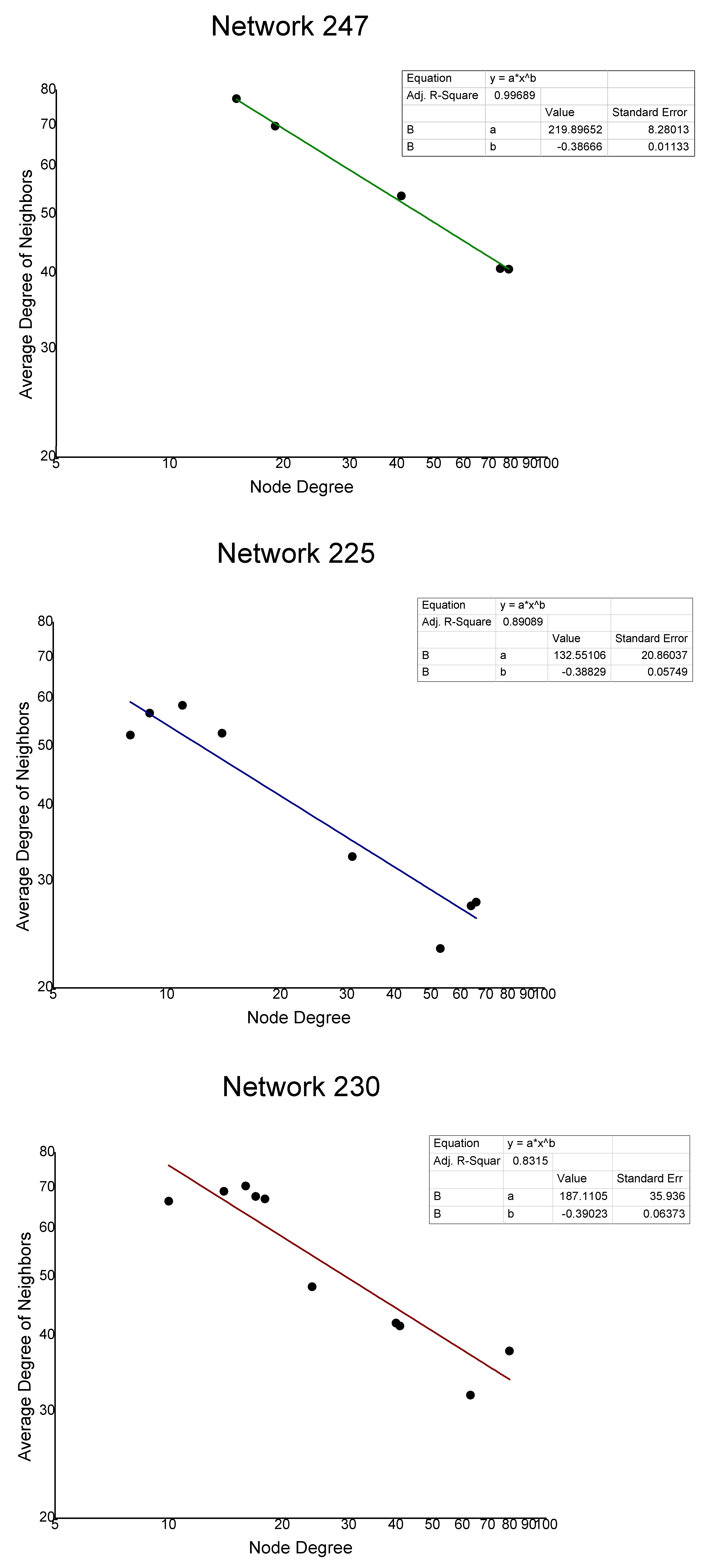}
\caption{Node degree vs. the average degree of neighbor nodes in a log-log plot for the three networks. The slope of the line that fits the data gives the assortativity values. The three networks have similar and negative assortativity values. }
\label{assortativity}
\end{figure}

The study of the degree distribution of the links did not show particular trends in any of the three networks. So it cannot be said that they have a scale-free structure. On the other hand, all three networks have a disassortative nature, as expected for food networks (\cite{Newman-MEJ-2002, Newman-MEJ-2003, Stouffer}): nodes with many links are mostly connected with nodes with a low number of links. In this calculation, the total number of links, given by the sum of the incoming and outgoing links, was considered for each node. The results are shown in Figure~\ref{assortativity} in which is plotted the node degree vs. the average degree of neighbor nodes for the three networks. The assortativity values are deduced from the slope of the lines that fitted the data in the log-log plot. All three networks have similar and negative assortativity values: $-0.387\pm 0.011$, $-0.388\pm 0.057$ and $-0.390\pm 0.064$ for networks 247, 225 and 230 respectively.

\subsection{Robustness of the three food webs}
\begin{figure}[tbp]
\centering
\includegraphics[width=1\columnwidth]{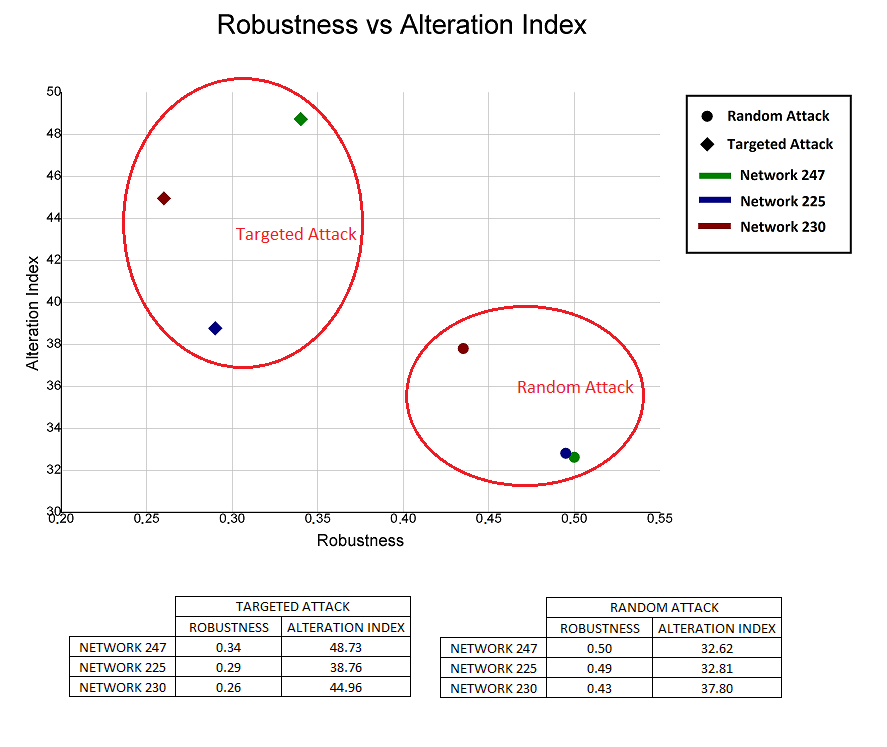}
\caption{Above are shown the values of the alteration index as a function of the robustness that the three networks have shown if subjected to targeted and random attacks. The tables below show the robustness and the alteration index for the three networks, in the case of targeted attacks and random attacks. The values for random attacks are the average values obtained on 10 tests.}
\label{robustness}
\end{figure}

We calculate the value of the robustness of the networks according to the definition provided by \cite{Dunne-et-al-2002b}. Using the  model described in \cite{Conti}, random and targeted attacks, to the species with the highest closeness centrality (for the same values of the closeness centrality it has been chosen to remove the species that is most abundant), were simulated. \\
We proceed in the following way: starting from the undisturbed system, the closeness centrality is calculated for all nodes and the one with the highest value is selected and removed. The dynamics of the system and the possible occurrence of secondary extinctions are therefore observed. Once the system has reached a condition of stability, the closeness centrality values are recalculated for all nodes and once again the one with the highest value is selected and removed. This process ends when half of the species have disappeared from the ecosystem (both because of removals and because of secondary extinctions). The robustness values for the three networks in the case of targeted attacks and in the case of random attacks (averaged over 10 differents simulations), together with the alteration index values, are shown in Figure~\ref{robustness}.\\
From the results obtained, some observations can be made. The first concerns the comparison between networks. Network 247 proves to be the most robust both in the case of targeted attacks and in the case of random attacks, confirming what already suggested by the values of the clustering coefficient and of the connectance; network 225 follows and finally network 230. The second observation concerns a comparison between types of attacks: all three networks are more robust against random attacks rather than targeted attacks. The last consideration concerns the relation between the alteration index and the robustness of the system. The alteration suffered by the ecosystem depends on the robustness of the network. The alteration index proves to be a good parameter for measuring the disgregation of the system. As expected, in the tests carried out, when robustness is greater, the alteration index is smaller, with the only exception of the case in wich network 247 is subject to targeted attack. In the next paragraph we try to explain this apparent contradiction.

\subsection{Trend of alteration index, connectance, complexity and species richness}
\begin{figure}[tbp]
\centering
\includegraphics[width=0.6\columnwidth]{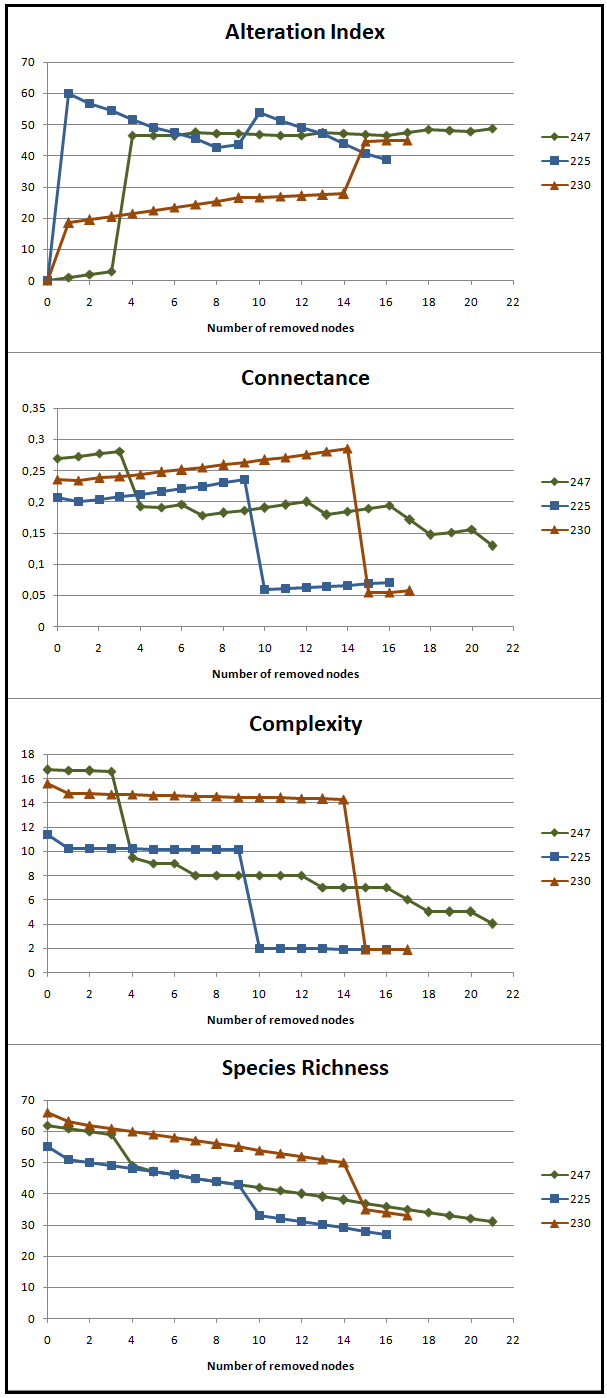}
\caption{Trend of the alteration index, the connectance, the complexity and species richness, as the nodes with the highest closeness centrality are removed, for the three food webs.}
\label{AI-C-C-B}
\end{figure}

Figure~\ref{AI-C-C-B} shows the trend of the alteration index, the connectance, the complexity and species richness, as the nodes with the highest closeness centrality are removed for the calculation of the robustness against targeted attacks.
We note that connectance and complexity have very similar trends: these decrease much more gradually and ultimately have higher values for network 247 than for the other two networks.
The most robust network (web 247) is therefore the one that has higher values of connectance and complexity both at the beginning and at the end of the simulation.
From the trend of species richness we can see the moments in which secondary extinctions occurred. Note that the major secondary extinctions occurred earlier in the 247 network than in the other two networks. This explains the evident increase in the alteration index in network 247 which is larger than that recorded for the other two networks: at the time of secondary extinctions, network 247 has a greater number of nodes than webs 225 and 230 and, as a consequence, the number of species that undergo alteration is greater (remember that the alteration index is additive with respect to the abundance variations suffered by the species present in the system).
This also explains why the alteration index of networks 225 and 230 undergoes a greater increase at the beginning, when only a small fraction of secondary extinctions occurs, and less when secondary extinctions are more: at the time of the major secondary extinctions, there are fewer species and therefore fewer species undergo an alteration.

\section{Conclusions}
In this study three soil ecosystems, that differ in the type of management and in the concentration of some elements, have been compared in the attempt to understand if and how anthropogenic action affects soil ecosystems.\\
With regard to the topology, it has been found that the structure of all three networks is small world. Furthermore, all three networks have a disassortative nature as expected for food webs (\cite{Newman-MEJ-2002, Newman-MEJ-2003, Stouffer}). Differences were found in the values of the clustering coefficient, of the connectance and of the complexity. These values are greater for network 247 and this suggested a greater robustness of this network than the other two (\cite{Dunne-et-al-2002b}). The calculation of the robustness, through a dynamic model (\cite{Conti}), confirmed this hypothesis. It therefore appears that the ecosystem related to the fallowed pastures with low pressure management is more robust than the two ecosystems  releted to organic farms subject to middle intensity management.\\
There are also differences between the two sites dedicated to organic agriculture, that could be connected to the different concentration of elements present in the soil and therefore to the anthropic action. The value of the clustering coefficient and the calculation of robustness suggest that network 225 is more robust than network 230, the site with high concentrations of heavy metals in the soil. This result is in contrast with what \cite{Dunne-et-al-2002b} affirmed, about the positive relationship between connectance and robusteness, but remaining available networks may constitute a larger sample on which to test this correspondence.\\
It is quite intuitive to say that the nutrients of the soil and its composition affect the resources available to soil organisms. This affects the type and abundance of organisms present in the soil and therefore the structure of the network from the lowest to the highest trophic levels of the food web (\cite{Wall, Clay, Powell}). Despite this, in order to fully ascertain the correlations existing between the robustness of the trophic network and the type of management to which the site is subjected as well as the type of soil composition, further studies would be required in order to shed light on the real impact that this type of management could have on the robustness of soil ecosystems.
 If the evidences suggested by this study were to be confirmed, the robustness shown by the networks could be useful for evaluating, from an ecological point of view, the sustainability of the agricultural practices to which the ecosystem is subject.

\section*{References}

\bibliography{mybibfile}

\end{document}